\documentclass[aps,preprint,nofootinbib,floatfix]{revtex4}

\usepackage{color}

\pdfoutput=1
\usepackage{graphicx}
\usepackage[latin1]{inputenc}
\usepackage{amsmath,amssymb}
\usepackage{hyperref}
\usepackage{epstopdf}
\usepackage{slashed}

\newcommand{\lsim}{\mathrel{\mathop{\kern 0pt \rlap
  {\raise.2ex\hbox{$<$}}}
  \lower.9ex\hbox{\kern-.190em $\sim$}}}
\newcommand{\gsim}{\mathrel{\mathop{\kern 0pt \rlap
  {\raise.2ex\hbox{$>$}}}
  \lower.9ex\hbox{\kern-.190em $\sim$}}}

\def  \bcen   {\begin{center}}
\def  \ecen   {\end{center}}
\def  \beq    {\begin{equation}}
\def  \eeq    {\end{equation}}
\def  \bpm    {\begin{pmatrix}}
\def  \epm    {\end{pmatrix}}
\def  \beqa   {\begin{eqnarray}}
\def  \eeqa   {\end{eqnarray}}
\def  \nn     {\nonumber }
\def\bea{\begin{eqnarray}}
\def\eea{\end{eqnarray}}

\def\ga   {\gamma}

\def\th   {\theta}

\def\La   {\Lambda}

\def\sig   {\sigma}

\def\nn{\nonumber}
\def\lee { \left( }
\def\rii { \right) }

\def\bnu {\bar{\nu}}
\def\bmu {\bar{\mu}}
\def\eps {\epsilon}

\begin{document}

{\small
\begin{flushright}
CP3-Origins-2018-015 DNRF90 \\
DO-TH 18/10 \\
\end{flushright} }

\title{Polarized gamma rays from dark matter annihilations}
\author{Wei-Chih Huang$^{1,2}$}
\author{Kin-Wang Ng$^{3,4}$}

\affiliation{
\small{
$^1$CP$^3$ Origins, University of Southern Denmark, Campusvej 55, DK-5230 Odense M, Denmark\\
$^2$Fakult\"at f\"ur Physik, Technische Universit\"at Dortmund,
44221 Dortmund, Germany \\
$^3$Institute of Physics, Academia Sinica, Taipei 11529, Taiwan\\
$^4$Institute of Astronomy and Astrophysics, Academia Sinica, Taipei 11529, Taiwan
}
}


\begin{abstract}
In this paper, we explore the possibility of a linearly polarized gamma-ray signal from dark matter annihilations in the Galactic center. Considering neutral weakly interacting massive particles, a polarized gamma-ray signal
can be realized by a two-component dark matter model of Majorana fermions with an anapole moment. We discuss the spin alignment of such dark matter fermions in the Galactic center and then estimate the intensity and the polarizability of the final-state electromagnetic radiation in the dark matter annihilations. For low-mass dark matter, the photon flux at sub-GeV energies may be polarized at a level detectable in current X-ray polarimeters.
Depending on the mass ratio between the final-state fermion and DM, the degree of polarization at the mass threshold can reach $70\%$ or even higher,
providing us with a new tool for probing the nature of dark matter in future gamma-ray polarization experiments.

\end{abstract}

\maketitle

\section{Introduction \label{section:1}}
Recent cosmological observations concordantly predict a spatially flat universe with $5\%$ baryons, $25\%$ cold dark matter (CDM),
and $70\%$ vacuum-like dark energy~\cite{Ade:2015xua}.
Unveiling the mystery of the dark components is one of the most important problems in science.
Although the nature of CDM is yet unknown, it has been successfully considered as elementary weakly interacting particles (WIMPs) 
with regard to the relic abundance and the formation of cosmic structures. Well-motivated candidates,
such as the lightest supersymmetric particle, extra dimension, hidden sector, and Higgs portal DM, 
 have long been sought after in experimental direct and indirect searches as well as at colliders.  So far all searches for WIMPs remain elusive, 
giving us stringent constraints on the scattering cross-sections of WIMPs with Standard Model~(SM) particles (see Ref.~\cite{Olive:2016xmw} for a recent review).

In the direct-detection experiment, presumably WIMPs in the Galactic dark CDM halo scatter with target nuclei in the detector that measures the recoil
energy of the nuclei, with background contamination mostly removed by the signal discrimination method. The indirect detection of WIMPs accumulated in the
solar core or Galactic center~(GC) is to search for signals coming from their decay or annihilation products such as gamma~($\ga$) rays, positions, antiprotons, and neutrinos.
This has been proven workable and is complementary to the direct detection; however, the observation is often masked by uncertain astrophysical background.
The removal of the astrophysical background is a challenging problem, so any characteristic feature of an indirect signal will be very useful for us 
to distinguish between WIMPs and astrophysical background sources. Most studies of the indirect signals have concentrated on the spectral fluxes of $\ga$ rays, positions, and neutrinos, whereas the polarization of $\ga$ rays has been scarcely discussed. 
Since future high-energy $\ga$-ray detectors are equipped with sensitive polarization capability~\cite{Aharonian:2017xup}, 
we aim to study the polarization of $\ga$ rays from WIMP annihilations in the GC that may enable us to separate the genuine signal from the astrophysical $\ga$-ray background. Here we are concerned with linear polarization. The possibilities for a net circular polarization of $\ga$ rays from DM annihilations or decays
have been explored in Refs.~\cite{Ibarra:2016fco,Kumar:2016cum,Gorbunov:2016zxf,Boehm:2017nrl,Elagin:2017cgu}.

\section{Polarized DM \label{section:2}}
We consider neutral WIMPs that annihilate into SM particles. In order to have polarized $\ga$ rays in the annihilation products, WIMPs must carry spins that can be aligned by an external directional field.
One of possible DM models is Majorana fermions\footnote{For Dirac DM, particles and antiparticles will be polarized along opposite directions and thus there is no preferred direction for DM annihilations.} with an anapole moment. In fact, such a WIMP called anapole DM has been proposed and studied~\cite{Pospelov:2000bq,Ho:2012bg,Fitzpatrick:2010br,Frandsen:2013cna,Gresham:2013mua,DelNobile:2014eta} as a kind of DM that interacts with ordinary matter via a spin-current electromagnetic interaction. The interaction Hamiltonian in the non-relativistic limit is given by
\begin{equation}
H_I=-\frac{g}{\Lambda^2}{\vec\sigma}\cdot{\vec J},
\label{EMint}
\end{equation}
where $g$ is a coupling constant, $\Lambda$ is an energy scale, $\vec\sigma$ are the Pauli spin matrices, and ${\vec J}={\vec\nabla}\times{\vec B}$ is the electromagnetic current density. 

Observation of stellar orbits in the GC has indicated that a supermassive black hole resides at the center of the Galaxy~\cite{Gillessen:2008qv}. Collective electric currents associated with gas accretion onto the black hole create large-scale magnetic fields~\cite{Shapiro:1983du}. The electric current density can be estimated as $J\sim B/h$, where $B$ is the magnetic field strength and $h$ is the thickness of the accretion disk. Then, the electromagnetic interaction energy~(\ref{EMint}) is of order
\begin{equation}
E_I\sim 10^{-29} {\rm eV} \,g\, \left(\frac{\rm GeV}{\Lambda}\right)^2 \left(\frac{B}{\rm Gauss}\right) \left(\frac{r_s}{h}\right) \left(\frac{M_{\odot}}{M}\right),
\end{equation}
where $M$ is the mass of the black hole and $r_s\equiv 2GM$ is the Schwarzschild radius. Here we use $M=10^6 M_{\odot}$.
The physical parameters in the accretion disk largely depend on the geometry and the temperature $T$ of the disk~\cite{Shapiro:1983du}. For a thin and cold disk ($h\ll r_s$ and $T < 0.1\, {\rm keV}$), the gas energy density is typically about $1\,{\rm g\, cm^{-3}}$, whereas the gas is expanded to a density of about $10^{-10}\,{\rm g\, cm^{-3}}$ in a thick and hot disk ($h\lesssim r_s$ and $T <  0.1 \, {\rm GeV}$ ). For equipartition magnetic fields,  $B \sim 10^{11} \,{\rm Gauss}$ and $\sim 10^6\, {\rm Gauss}$ are in the thin-cold and the thick-hot disk, respectively. Assuming $g=1$ and $\Lambda=100\, {\rm GeV}$, the interaction energy is at least $E_I \sim 10^{-33} {\rm eV}$. For a thin-cold disk, $E_I \gg 10^{-28} {\rm eV}$. 

In the presence of the directional current, the degree of spin alignment of DM particles along the current flow is governed by the Boltzmann factor, $e^{-E_I/T_s}$, where $T_s$ is the spin temperature. It is difficult to determine $T_s$, which would depend on the history of the DM halo formation and the baryonic environment. The dominant process for DM spin flips in the GC is the DM-proton scattering $D \, p \to D \, p$. Direct searches severely constrain the DM-proton cross-section $ \sigma_{D \, p \to D \, p} $, which is spin-independent but velocity-suppressed in the case of anapole interactions. However, the bound becomes much less stringent for DM masses below 5 GeV or so -- for 1 GeV DM, the DM-nucleon cross-section can be as large as $10^{-38}$ cm$^2$~(see, for instance, Fig.~6 in Ref.~\cite{Petricca:2017zdp})~\footnote{
Note that in the low-mass region, LHC searches using events with large missing transverse momentum and one or more energetic jets~\cite{CMS:2016pod}, 
in the context of simplified models,  set limits on the cross-section: $ \sigma_{D \, p \to D \, p} \lesssim 10^{-43}$ cm$^2$ for the vector and axial-vector interactions.
The LHC bounds in general do not apply to anapole interactions as they are usually loop-induced and hence there is no resonance enhancement from the mediator as in the simplified models.}.
In a thick-hot disk, the proton number density in the GC is about $n_p \sim 10^{14} \, {\rm cm^{-3}}$ and the proton velocity is $v_p\sim 0.1c$. In the following, we assume a typical DM velocity in the DM halo, $v_D\sim 10^{-3}c$. The typical timescale for the scattering process is given by $\tau=(n_p \sigma_{D \, p \to D \, p} v)^{-1} \sim 10^{8} {\rm \, yr}$, where the relative velocity is $v\sim v_p\sim 0.1c$, which is smaller than the age of the Galaxy of order $10^{10} {\rm \, yr}$. As such, one would expect that the DM spins are random. However,
in a thicker ($h > r_s$) and/or less hot ($T \ll 0.1 \, {\rm GeV}$) disk, the scattering timescale may easily exceed $10^{10} {\rm \, yr}$.  In this situation, if the initial $T_s$ is smaller than $E_I$, DM will stay in the lower energy states with spins lining up with the current.

On the other hand, the thin-cold disk has $n_p \sim 10^{24} \, {\rm cm^{-3}}$ and $v_p\sim 10^{-4}c$. Hence we have $v\sim v_D\sim 10^{-3}c$ and that $\tau \sim 10 {\rm \, yr}$, which is much shorter than the Galactic age. As a result of multiple DM spin flipping in the DM-proton scatterings, one would expect that DM spins are randomized and thus that $T_s \gg E_I$. However, this argument is incomplete under the consideration of the principle of minimum energy, which suggests that DM particles will relax to the ground states with spins lining up with the external current. In fact, the mechanism responsible for the energy minimization is the bremsstrahlung cooling process, $D^* p \to D \, p \, \ga$, where $D^{(*)}$ is the ground~(excited) state DM particle and $\ga$ emitted from the proton is the bremsstrahlung photon that carries away the DM kinetic energy and the excitation energy $E_I$. The cross-section can be estimated~\cite{Birkedal:2005ep} from the DM-proton scattering process as $(\alpha /\pi) \sigma_{D \, p \to D \, p} $.  The transition time for DM from the excited to ground state, that is inversely proportional to $(\alpha /\pi) \sigma_{D \, p \to D \, p}  n_p v$, is $\tau \sim 10^3$\, yr, which is again much shorter than the Galactic age . As long as the incoming particles have kinetic energy larger than the excitation energy, the same inverse process $D \, p \to D^* p \, \ga$ is efficient enough to reverse the spin. In the beginning,  DM spins are randomized while their kinetic energy is being dissipated away by bremsstrahlung photons. During the course of the Galactic lifetime, most DM kinetic energy is lost through the bremsstrahlung cooling and subsequently a fraction of the DM is slowed down or even stopped. These slow-moving DM particles do not interact with colder gas in the outer region of the disk and eventually de-excited to the ground state by emitting virtual photons to the current background. These virtual photons are absorbed in the current by creating small perturbation in the current background of waveform, $e^{-iE_I t\pm i {\vec q}\cdot{\vec x}}$, where ${\vec q}$ is the DM momentum transfer, and the scattering amplitude is proportional to ${\tilde J}({\vec q})$ that is the Fourier transform of the current $J({\vec x})$. A detailed consideration of the cooling process and the de-excitation will be needed to assess the fraction of the polarized DM particles.

The above estimates, though somewhat contrived, have suggested that the anapole WIMPs may be polarized along the current flow on the accretion disk in the GC if initial DM spin temperatures are very low in some thick-hot disks. If the disk is thin and cold, a partial DM polarization may be possible. More detailed investigations should be in order, focusing on the formation of the DM halo and the accretion disk as well as the interaction between these two structures. This may help determining the DM spin temperature in the DM halo core as well as assessing the amount of DM spin alignment with the electric current in the disk. 

The main interest of the present work is to propose for the first time a possibility of a linearly polarized $\ga$-ray signal from DM annihilations. 
Here we have restricted ourselves to a neutral WIMP scenario, simply assuming that the WIMPs are Majorana fermions whose anapole moment allows them to be polarized in the Galactic core. Then, we consider the linear polarization of the $\ga$ rays from these DM annihilations. However, it would be interesting to explore other scenarios, for instance, by considering a dark mirror universe in which DM particles carrying dark magnetic dipole moment are polarized in an external dark magnetic field in the Galactic core. There exist many particle models for a dark mirror universe that contains dark photon; see, for instance, Refs.~\cite{Foot:1991bp,Berezhiani:1995am,Berezhiani:2000gw,Chacko:2005pe,Huang:2015wts}. Therefore, it may not be impossible to realize in some models on what we have suggested. Overall, the capability of detecting linear polarization in future $\ga$-ray observations will open a new window for us to look for DM and thus theoretical endeavors for a polarized $\ga$-ray signal should be warranted.  

\section{phenomenological model  \label{section:model}}
\begin{figure}[htp!]
\centering
\includegraphics[clip,width=0.5\linewidth]{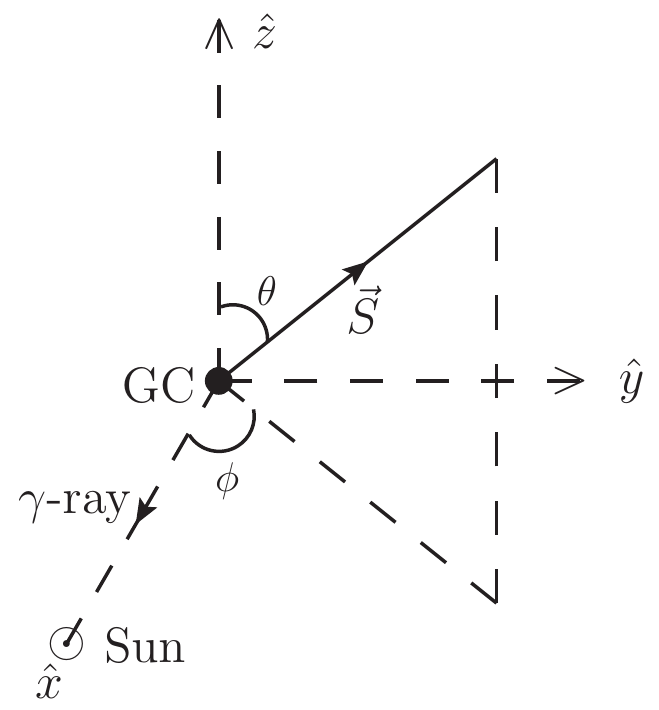}
\caption{The Galactic coordinate used in this work, where the sun lies on the $x$-axis and the DM polarization $\vec{S}$ is characterized by the polar and azimuthal angles, $\th$ and $\phi$, respectively.  }
\label{fig:DM_spin}  
\end{figure}

In this phenomenological model, we have two types of Majorana DM particles $\chi_1$ and $\chi_2$
of the nearly equal mass $m_\chi$.
The effective Lagrangian for the anapole interaction in terms of the 4-component spinor notation reads
\begin{align}
\mathcal{L} \supset  \lee \frac{g}{\La_1^2} \bar{\psi}_1  \ga^\mu\ga^5  \psi_1 
+ \frac{g}{\La_2^2}\bar{\psi}_2 \ga^\mu\ga^5  \psi_2
+ \frac{g}{\La_3^2}\bar{\psi}_1 \lee  i \ga^\mu + A\ga^\mu\ga^5 \rii \psi_2 \rii 
\partial^\nu F_{\mu\nu} \, ,
\label{eq:Lagr}
\end{align}
where
\begin{align}
\psi_1 =
\begin{pmatrix}
\chi_1 \\  \chi^\dag_1
\end{pmatrix} \,\, , \,\,
\psi_2 =
\begin{pmatrix}
\chi_2 \\  \chi^\dag_2
\end{pmatrix}
\label{eq:4-2_conv}
\end{align} 
with $\chi_1$ and $\chi_2$ being two-component Weyl spinors\footnote{We here follow the notations used in Ref.~\cite{Dreiner:2008tw}.}. 
Interactions such as $\bar{\psi}_i \ga^\mu \psi_i$ do {\it not} exist since $\chi_1$ and $\chi_2$ are Majorana particles. 
Note that $\bar{\psi}_1 i \ga^\mu \psi_2$ is Hermitian~(real) which can be proven by the following equality:
\begin{align}
\chi_i \sig^\mu \chi^\dag_j = - \chi^\dag_j \bar{\sig}^\mu \chi_i .
\end{align}
It is clear that $\chi_1$ and $\chi_2$ can be polarized along the same direction through electromagnetic anapole interactions, given
a strong electric current in the GC, while the mixing term gives rise to the transition between $\chi_1$ and $\chi_2$, resulting
in the equal amount of $\chi_1$ and $\chi_2$ if the transition is fast enough.
Throughout this work, we assume that $\chi_1$ and $\chi_2$ are polarized along the same direction $\vec{S}$,
denoted by the azimuthal and polar angle, $\phi$ and $\theta$ respectively, as shown in Fig.~\ref{fig:DM_spin} where we set the connection between GC and the sun
to be the $x$-axis. 

\begin{figure}[htp!]
\centering
\includegraphics[clip,width=0.46\linewidth]{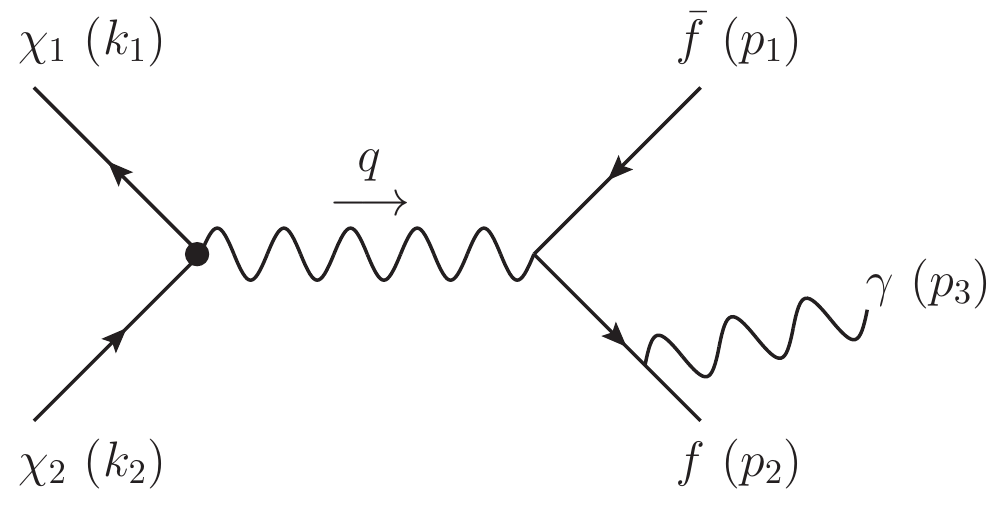}
\includegraphics[clip,width=0.46\linewidth]{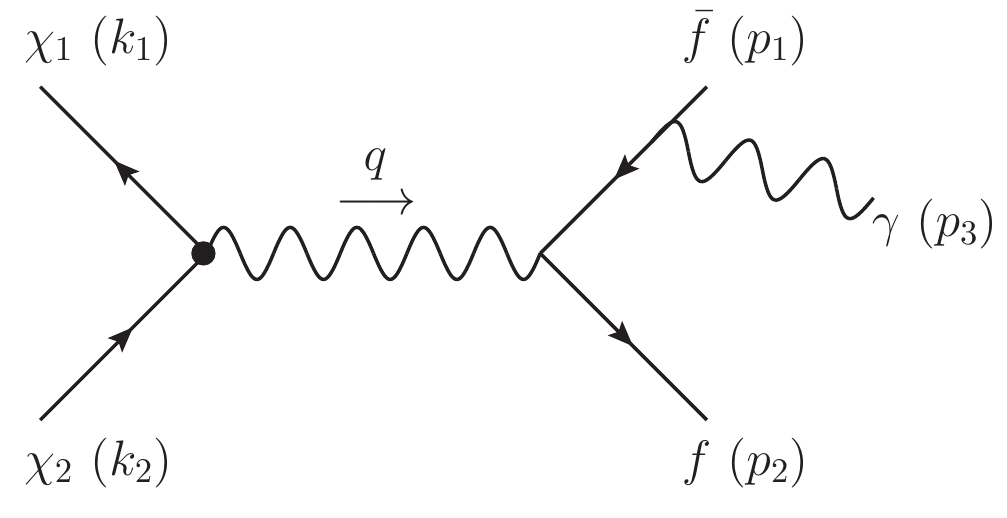}
\caption{Feynman diagrams for the dominant DM annihilation process $\chi_1 \chi_2 \to \bar{f} f$ with final state radiation.}
\label{fig:DM_ff}  
\end{figure}

Due to the facts that we intend to study $\ga$-ray polarizations from annihilations of polarized $\chi_1$ and $\chi_2$ into SM fermions~($f$ and $\bar{f}$) followed by the final state radiation as shown in Fig.~\ref{fig:DM_ff} and that the current DM velocity is
very low $v \sim 10^{-3} c$,
relevant annihilation processes should be independent of the DM velocity, i.e, $s$-wave~(total angular momentum of the DM system is zero, $L=0$).
For $\chi_{(1,2)} \chi_{(1,2)} \to \ga \to \bar{f} f $, only one term in Eq.~\eqref{eq:Lagr}, $\bar{\psi}_1 \ga^\mu \psi_2$, has a component of $L=0$
with a total spin $S=1$~\cite{Kumar:2013iva}. By contrast,
terms involving $\ga^5$ do have a $s$-wave component but with $S=0$, and so there is no preferred direction for outgoing photon polarizations in this case. 
All in all, we  consider only the dominant process  $\chi_1 \chi_2 \to \bar{f}f$
induced by the operator $\bar{\psi}_1 \ga^\mu \psi_2$. That is the reason why two Majorana DM $\chi_1$ and $\chi_2$
are required to create the polarized photon. As mentioned above, a Dirac DM candidate will not work as a particle will be polarized in an opposite direction to an antiparticle and thus there is no favored direction in particle-antiparticle annihilation.
 
Employing the Feynman rules for two-component Weyl spinors~\cite{Dreiner:2008tw},
on the DM side the amplitude of $\chi_1-\chi_2$ annihilation into a SM fermion pair with a prompt photon emission is,
in the limit of zero DM velocity,
\begin{align}
i \mathcal{M}_1^\mu \lee \bar{\psi}_1 \ga^\mu \psi_2 \rii = \frac{i g}{\La^2_3}
2m_\chi \lee 0, -\cos\th\cos\phi + i\sin\phi,  -\cos\th\sin\phi - i \cos\phi, \sin\th \rii q^2, 
\end{align}    
where $q^2$, the transferred momentum squared, comes from $\partial^\nu \partial_\nu  A_\mu$
and will be canceled by the denominator of the photon propagator. The other contribution from $\partial^\nu \partial_\mu  A_\nu$ vanishes
when contracted with $ \bar{\psi}_1 \ga^\mu \psi_2$, which can be understood by simply applying the equation of motion:
$ \bar{\psi}_1(k_1) (\slashed{k_1} + \slashed{k_2}) \psi_2(k_2) \sim \bar{\psi}_1(k_1) (-m_\chi  + m_\chi) \psi_2(k_2)$.

The amplitude on the SM side has two contributions corresponding to the final photon attached to either of the outgoing fermions.      
It is straightforward to compute the amplitude squared and sum over final fermion spins. The result reads
\begin{align}
 \mathcal{M}_{2\mu} \mathcal{M}^*_{2{\bmu}} &= Q^2_f e^2 \Bigg(
 Tr \left[ \ga_\mu  \frac{\lee \slashed{p}_2 - \slashed{q} + m_f \rii}{\lee p_2 - q \rii^2  - m^2_f}
 \ga_\nu \lee \slashed{p}_1 -m_f \rii 
\ga_{\bnu}   \frac{\lee \slashed{p}_2 - \slashed{q} + m_f \rii}{\lee p_2 - q \rii^2  - m^2_f}
\ga_{\bmu} \lee \slashed{p}_2 + m_f \rii   \right]
\nn \\
&+ 
Tr \left[ \ga_\nu  \frac{\lee \slashed{q} - \slashed{p}_1 + m_f\rii}{\lee q - p_1 \rii^2 - m^2_f}
  \ga_\mu \lee \slashed{p}_1 -m_f \rii 
\ga_{\bmu}   \frac{\lee \slashed{q} - \slashed{p}_1 + m_f\rii}{\lee q - p_1 \rii^2 - m^2_f} 
\ga_{\bnu} \lee \slashed{p}_2 + m_f \rii   \right] \nn\\
&+
\text{mixing terms}  \Bigg) \eps^{*\nu}\lee p_3\rii \eps^{\bnu}\lee p_3\rii ,
\end{align}
where $p_1~(p_2)$ is the momentum of $\bar{f}~(f)$,
$e$ is the electric coupling, and $Q_f$ is the fermion electric charge while the mixing terms refer to the interference between two diagrams in Fig.~\ref{fig:DM_ff}.
The symbol $\epsilon^\mu$ is the photon polarization vector; for example, $\epsilon=(0,0,0,1)$ for polarization along $z$.
Including the photon propagator, the square of the total amplitude is simply,
\begin{align}
\vert M \vert^2 = \mathcal{M}_1^\mu \mathcal{M}^{*{\bmu}}_{1}  \mathcal{M}_{2\mu} \mathcal{M}^*_{2{\bmu}} \frac{1}{ \lee q^2 \rii^2},
\end{align}
and the corresponding differential cross-section times the DM velocity $v$ in the limit of $v \to 0$
 becomes\footnote{For the 3-body phase integral, see, for instance, Ref.~\cite{Olive:2016xmw}.}
\begin{align}
v \frac{d \sig}{d\Omega} = \frac{1}{\lee 2 \pi \rii^5 32 m^2_\chi } 
\int^{{E_3}_{max}}_{{E_3}_{min}} d E_3
\int^{{E_1}_{max}}_{{E_1}_{min}} d E_1
\int_0^{2 \pi} d \ga' \vert M \vert^2,
\label{eq:Omega_dif}
\end{align}
where $\ga'$ is the rotation angle of the final state system with respect to the photon direction and the solid angle $\Omega$
indicates that only the outgoing photons along the $x$-axis can reach the earth as displayed
in Fig.~\ref{fig:DM_spin}.
The bounds on the energy of $\bar{f}$, given the photon energy $E_3$ are
\begin{align}
{E_1}_{min} &=  m_\chi -\frac{E_3}{2}  - \frac{E_3\sqrt{ m_\chi \lee m_\chi - E_3\rii \lee m^2_\chi - m_\chi E_3 - m^2_f \rii }}
{2 m^2_\chi - 2m_\chi E_3 }  \nn\\
 {E_1}_{max} &= 2 m_\chi - E_3 - {E_1}_{min} .
\end{align} 
The minimum~(maximal) $E_1$ occurs when the positron is along~(against) the photon direction. 
On the other hand,  the upper bound on the photon energy $E_3$ is
\begin{align}
 {E_3}_{max} &= \frac{m^2_\chi - m^2_e}{m_\chi} ,
\end{align} 
while the minimal $E_3$ is determined by the detector threshold of interest\footnote{The cross-section in fact becomes
divergent at $E_3=0$ and will be regulated by the virtue photon exchange.
We, nonetheless, are not interested in outgoing photons with very low energies.}.
As the final expression for the differential annihilation cross-section in Eq.~\eqref{eq:Omega_dif}
is unbearably lengthy and not very informative, we 
present only numerical results in the following sections.

\section{Photon kinematical distributions and polarization rate\label{section:kinematics}}
In this section, we discuss kinematics of the photon differential distributions and the energy dependence of the
polarization rate.
To simplify the computation, we focus on DM annihilation in the GC where the DM density is highest
and assume that the initial DM system has a total spin of one along the positive $z$-direction, corresponding to $\th=\phi=0$. It is straightforward to generalize to an arbitrary polarization direction.

\begin{figure}[htp!]
\centering
\includegraphics[clip,width=0.308\linewidth]{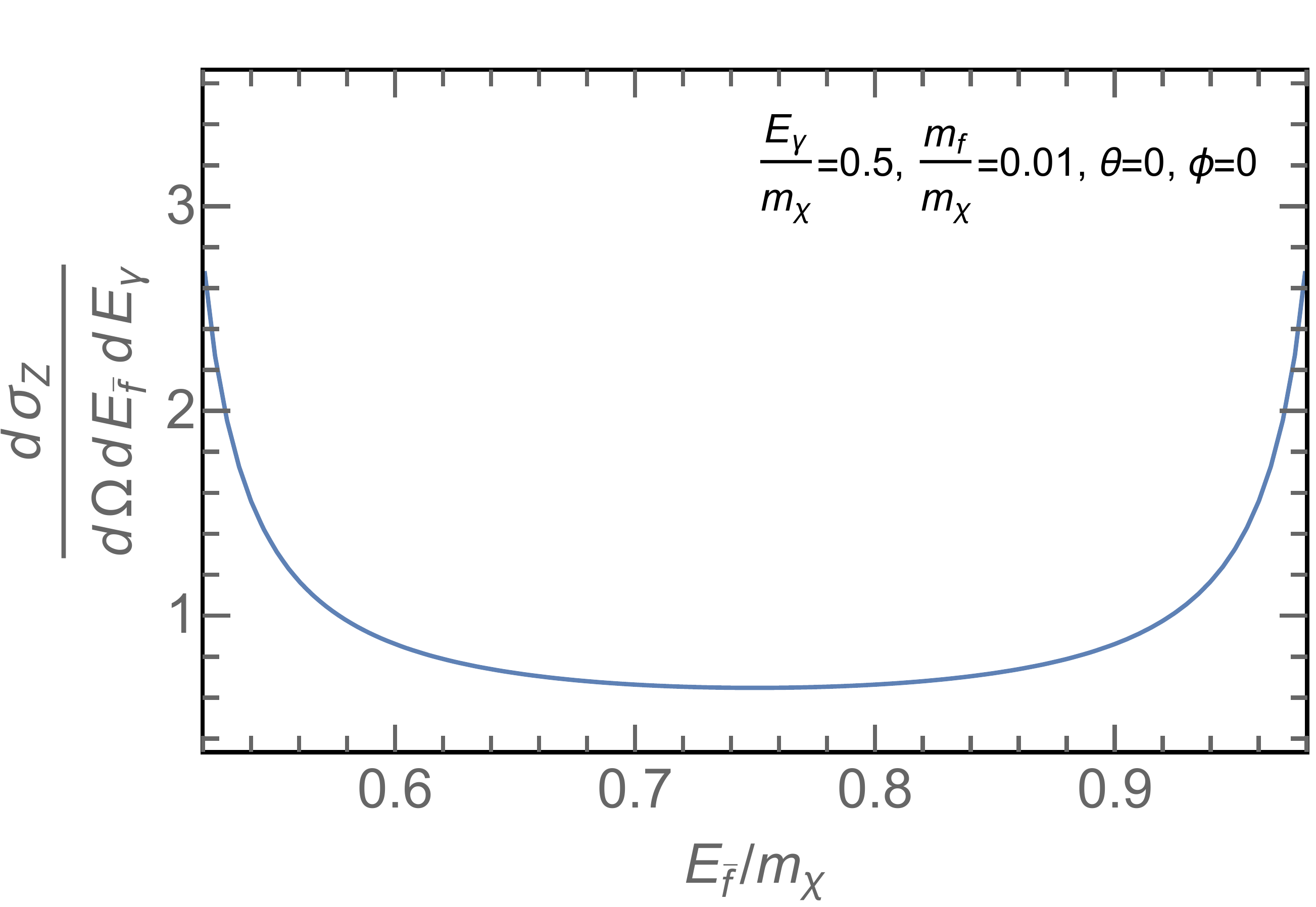}
\includegraphics[clip,width=0.33\linewidth]{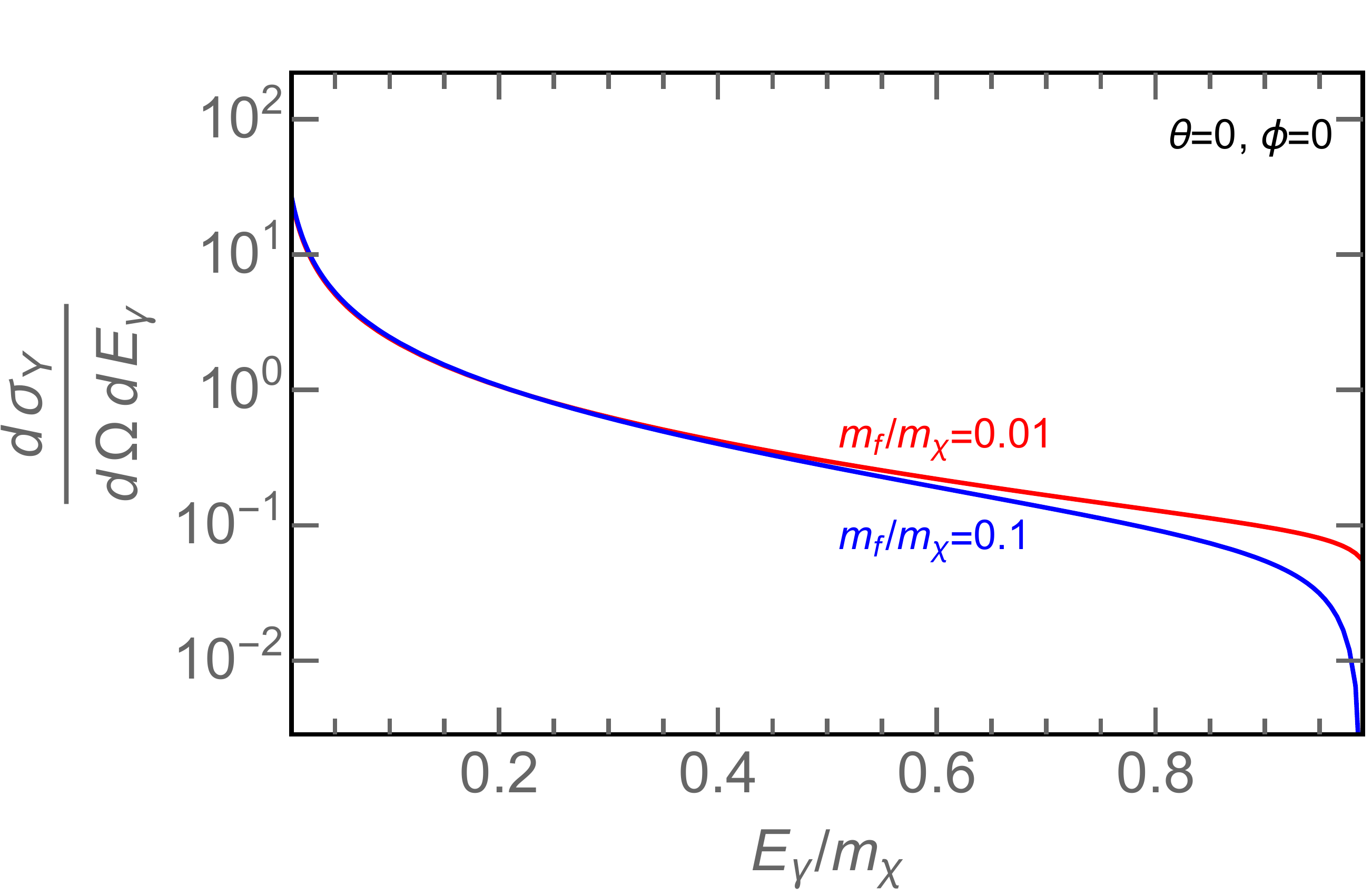}
 \includegraphics[clip,width=0.33\linewidth]{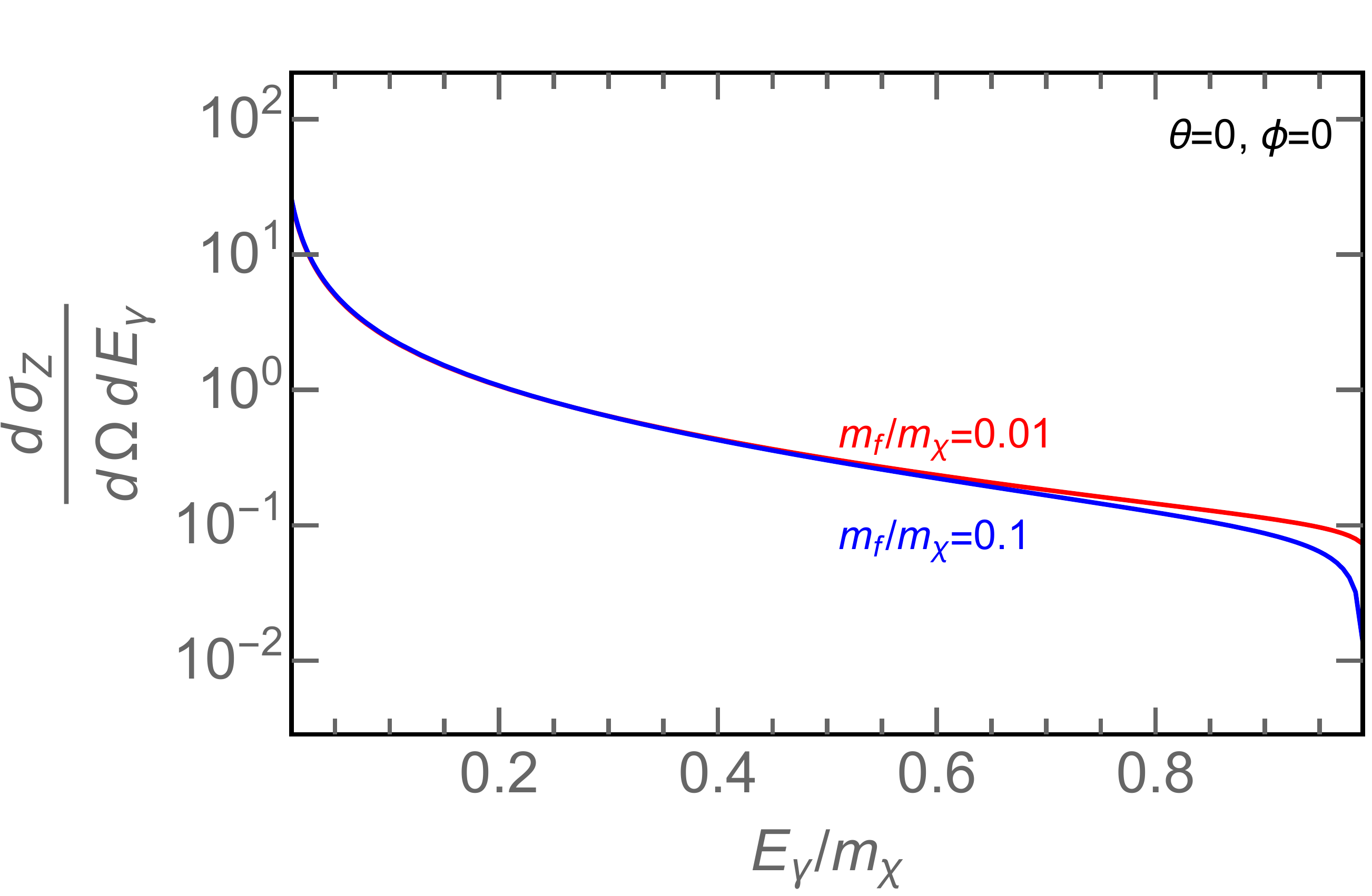}
\caption{Dependence of the differential annihilation cross-sections on $E_{\bar{f}}$~(left panel)
and $E_{\ga}$~(central and right panels), assuming that the DM system is polarized along $+ \hat{z}$.
The subscript $Z~(Y)$ of $\sigma$ refers to the $z$-polarized~($y$-polarized) photon.}
\label{fig:dsigma}  
\end{figure} 

We start with the differential annihilation cross-section as a function of $E_1$~(which is $E_{\bar{f}}$) and $E_3$~($E_{\ga}$)
displayed in Fig.~\ref{fig:dsigma}, where  the $y$-axis of all the panels is rescaled such that the total area below the curve is equal to unity.  
Given $E_\ga/m_{\chi}=0.5$ and $m_f/m_\chi=0.01$,
the left panel of Fig.~\ref{fig:dsigma} presents the differential cross-section for the $z$-polarized
photons as a function of $E_{\bar{f}}$. This demonstrates that
the differential cross-section are maximal when the photon is aligned with~(low $E_{\bar{f}}$) or against~(high $E_{\bar{f}}$)
the direction of $\bar{f}$. In other words, the maximum occurs when the photon is collinear with either $f$ or $\bar{f}$, in consistent with the statement in Ref.~\cite{Birkedal:2005ep}.
In the central panel~(for $y$-polarized photon)
and the right panel~(for $z$-polarized photon) of Fig.~\ref{fig:dsigma},
where $E_{\bar{f}}$ has been integrated over and the red~(blue) curve corresponds to $m_f/m_\chi = 0.01~(0.1)$,
 the photon energy spectrum peaks toward $E_\ga \to 0$ due to the infrared divergence.
The minimum of the differential cross-section occurs when both of $f$ and $\bar{f}$ are in the same direction but opposite
to that of the photon.
 
\begin{figure}[htp!]
\centering
\includegraphics[clip,width=0.5\linewidth]{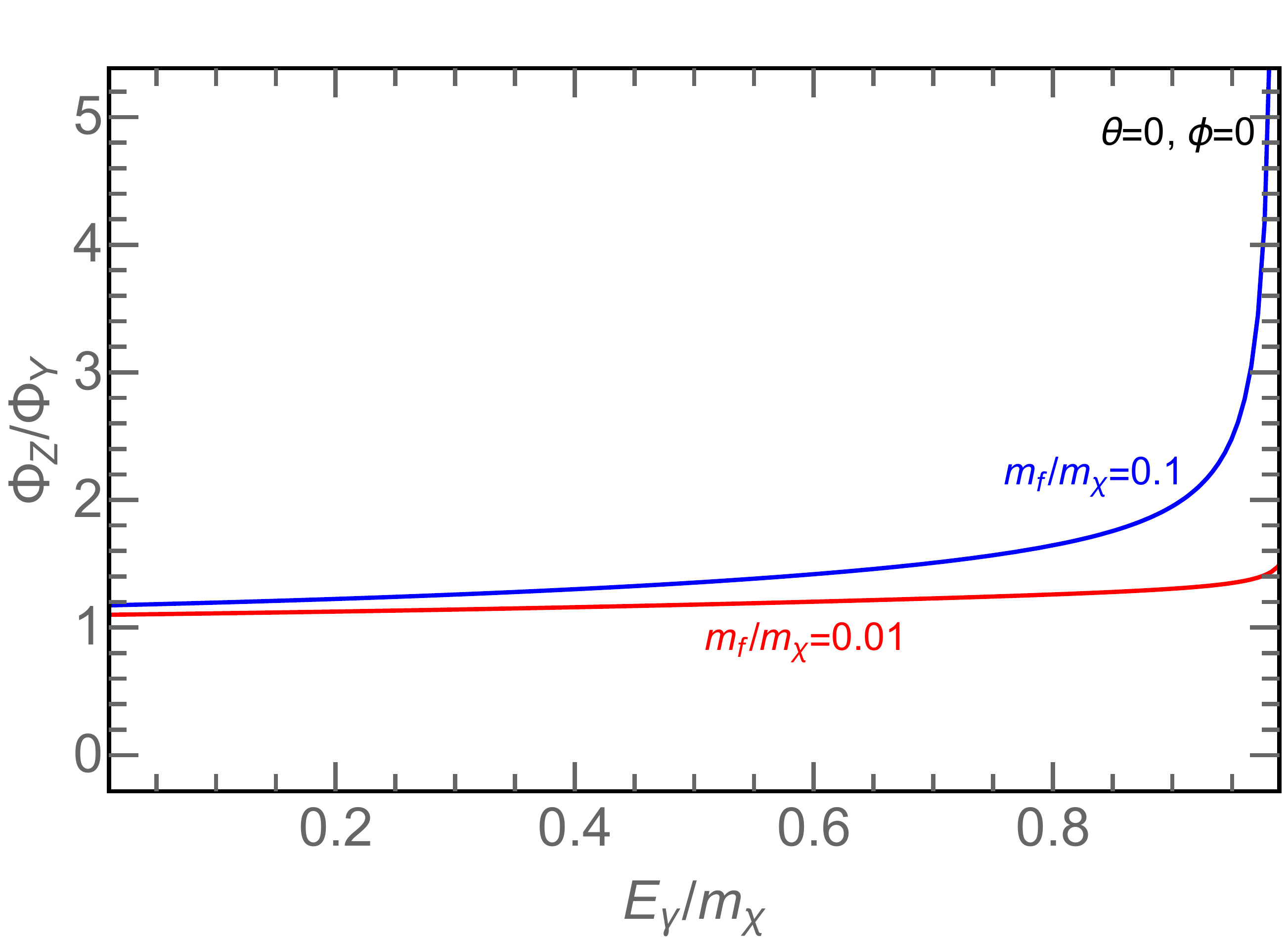}
\caption{The flux ratio of the $z$-polarized to $y$-polarized photon as a function of $E_{\ga}$,
assuming both $\chi_1$ and $\chi_2$ be polarized along $+\hat{z}$.}
\label{fig:Phi_ZY}  
\end{figure}

Fig.~\ref{fig:Phi_ZY} shows the ratio of the differential cross-section of the $z$-polarized photon
to that of the $y$-polarized one as a function of $E_\ga$, assuming that the DM particles $\chi_{1,2}$ are polarized along the positive $z$-direction.
This ratio is equivalent to the ratio of the corresponding photon fluxes $\Phi_Z/ \Phi_Y$.
We present two cases of different mass ratios: $m_f/m_\chi =0.1$~(blue line)
and $m_f/m_\chi =0.01$~(red line). It is clear that outgoing photons are more likely
to have the polarization along the $z$- than $y$-direction.

In general, the photons with larger energies are more likely to be polarized along the $z$-direction
than those with smaller energies. It can be understood in a naive argument based on angular momentum conservation as follows.
First, the electric current  couples a left-handed particle to a right-handed
anti-particle~(or a right-handed particle to a left-handed anti-particle).
Second, in the massless limit, the fermion chirality coincides with the helicity, which is the projection of the spin onto the direction of momentum.     
\begin{figure}[htp!]
\centering
\includegraphics[clip,width=0.5\linewidth]{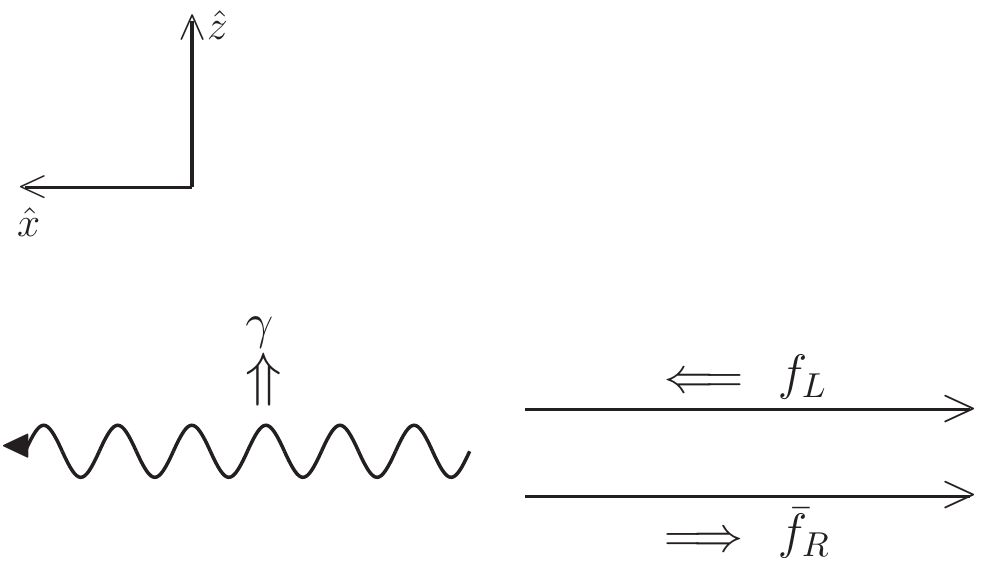}
\caption{The pictorial explanation of why $\Phi_Z/\Phi_Y$ increases as $E_\ga$ becomes larger.
For the energetic photon, the spins of the pair fermions cancel out so that the photon is polarized along the same direction as the total angular momentum of the initial DM system, i.e., $+\hat{z}$.}
\label{fig:L_con_aga}  
\end{figure} 
Third, for the high-energy photon with $f$ and $\bar{f}$ moving in the same direction,
the spin angular momenta of $f$ and $\bar{f}$ cancel each other such that the polarization of the photon has to be in the $z$-direction
to conserve the angular momentum as displayed in Fig.~\ref{fig:L_con_aga}, given that the initial DM system has $\vec{L}=0$ and $\vec{S}=+\hat{z}$.
Consequently, the high-energy photons tend to be polarized along the $z$-direction. By contrast, for the low energy photon,
the spins of the fermion pair add up to unity along the $x$-direction. In this case, a nonzero orbital angular momentum of the final states is required to conserve the total angular momentum, rendering $\Phi_Z/\Phi_Y$ smaller.    

\begin{figure}[htp!]
\centering
\includegraphics[clip,width=0.5\linewidth]{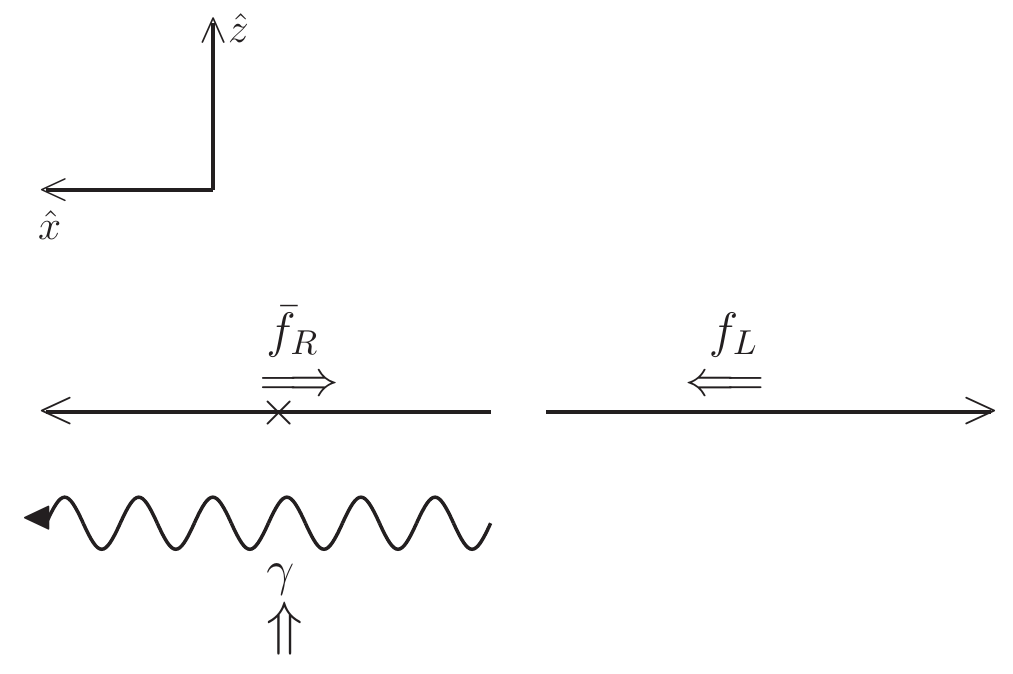}
\caption{The pictorial explanation of how a chirality flip~(a mass insertion) can increase the possibility of  
having the $z$-polarization for the low-energy photons.}
\label{fig:mass_ins}  
\end{figure}

On the other hand, it is noticeable that with a larger value of $m_f/m_\chi$ the photon polarization becomes more pronounced.
For the low-energy photons, it can be explained by a chirality flip due to the existence of the mass term as displayed
in Fig.~\ref{fig:mass_ins}. To be more concrete, because of the mass term, there exists a finite possibility that
even the right-handed $\bar{f}$ can have a left-handed helicity -- heavier mass, higher probability --
so that the spins of the fermion pair cancel each other, leading to the $z$-polarized photon.
On the other hand, for the high-energy region, with larger $m_f/m_\chi$
the $\Phi_Y$ decreases more dramatically than $\Phi_Z$ as the energy increases. This can be seen
by comparing the blue lines between the central and right panels of Fig.~\ref{fig:dsigma}. As a result, $\Phi_Z/\Phi_Y$ becomes much larger for $E_\ga/m_\chi$ close to unity.

\section{Estimation of Polarized photon flux} \label{section:flux}
As explained above, the only annihilation process with nonzero spin is induced
by the operator $\bar{\chi_1}\ga^\mu \chi_2 \partial^\nu F_{\mu\nu} $, which can be rewritten as
$\bar{\chi_1}\ga^\mu \chi_2 \bar{f} \ga_\mu f$.
The expected photon flux on Earth from DM annihilation  $\chi_1 \chi_2 \to \bar{f}f$ at the GC followed by
final state radiation (see Fig.~\ref{fig:DM_ff}) is~\footnote{For DM-induced photon flux computations, see, e.g., Ref.~\cite{Cirelli:2010xx} for more details.}
\begin{align}
\frac{d\Phi}{d\Omega dE_{\ga}} = \frac{r_\odot}{4} \lee \frac{\rho_\odot}{m_\chi} \rii^2 J
\sum_f\frac{d \sigma(\chi_1\chi_2 \to \bar{f}f\ga)}{d \Omega d E_{\ga}} v \, ,
\label{eq:ga_flux_dif}
\end{align}
where $\rho_\odot=0.3$ GeV/cm$^3$, $r_\odot=8.33$ kpc,
and $v$ is the relative velocity between $\chi_1$ and $\chi_2$,
while $J$ stands for the $J$ factor, corresponding to the DM density~(squared) integral along
the line of sight given a solid angle $d\Omega$:
\begin{align}
J= \int_{\text{l.o.s}} \frac{ds}{r_\odot} \lee \frac{\rho}{\rho_\odot}\rii^2 \, .
\end{align}
Here, we assume $\chi_1$ and $\chi_2$ have the same density and be polarized along $+ \hat{z}$.
This results in an anisotropic system. Then,
instead of the conventional assumption of
the isotropic photon distribution that leads to a factor of $4 \pi$ in the denominator,
$ d \sigma/d \Omega d E_{\ga}$ is employed in Eq.~\eqref{eq:ga_flux_dif} and
can be computed based on Eq.~\eqref{eq:Omega_dif}.

\begin{figure}[htp!]
\centering
\includegraphics[clip,width=0.5\linewidth]{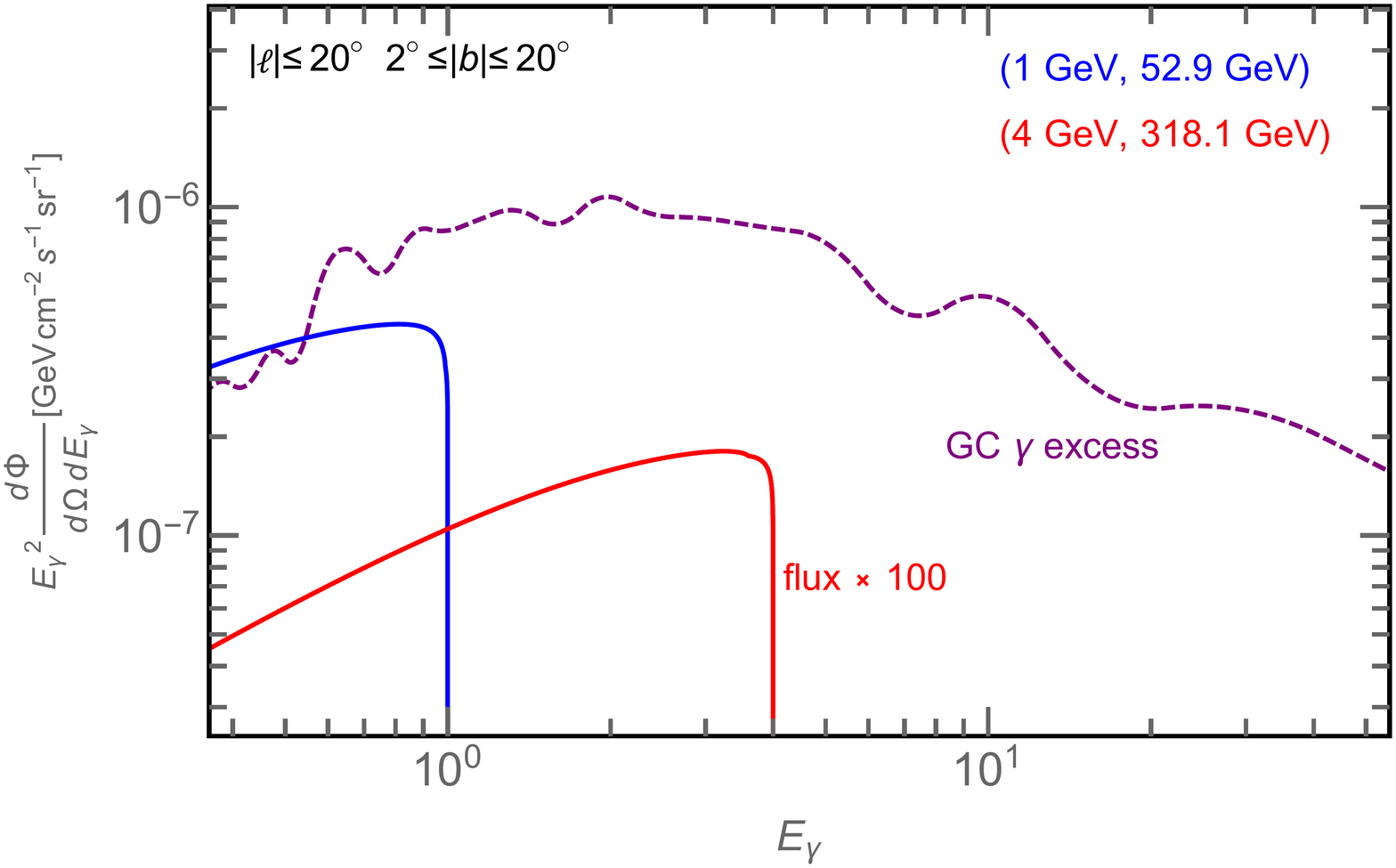}
\caption{$\ga$ flux from DM annihilation}
\label{fig:ga_flux}  
\end{figure}

On the other hand, the same operator will also give rise to sizable spin-independent DM-proton interactions
$\chi_{(1,2)} p \to \chi_{(2,1)} p$,
\begin{align}
\sig_{\text{DM-p}} = \frac{  e^4 \mu^2}{\pi \La^4_3} \, ,
\end{align}
 where $\mu$ is the reduced mass of the DM-proton system and the coupling constant $g$ in Eq.~\eqref{eq:Lagr}
 is set to the electric coupling $e$.
 For $m_{\chi}$ below 5 GeV or so, the bounds on the spin-independent DM-nucleon
cross-section become much weaker, implying a smaller $\La_3$ and hence a larger
photon flux from $\chi_1 - \chi_2$ annihilation.
In Fig.~\ref{fig:ga_flux}, we show the expected $\ga$-ray flux for 1 GeV~(blue solid line) and
4 GeV~(red solid line, which is the flux multiplied by a factor of 100), while the purple dashed line
represents the GC $\ga$-ray
excess taken from Ref.~\cite{Calore:2014xka} for the region of interest, 
$|\ell | \leq 20^\circ$ and $  2^\circ \leq | b | \leq 20^\circ$, which has
$J= 25.8$, assuming the Navarro-Frenk-White~(NFW) DM profile~\cite{Navarro:1995iw}.
The entries in the parentheses indicate the values of $m_\chi$ and $\La_3$ respectively, 
where   $\sig^{\text{SI}}_{\text{DM-nucleon}} \leq 9 \times 10^{-39}$~($1.7 \times 10^{-41}$) cm$^2$
for DM of 1~(4) GeV from the CRESST-II~\cite{Angloher:2015ewa}~(CDMSlite~\cite{Agnese:2015nto}) experiment are used.
Clearly, DM of 4 GeV is more constrained by direct detection than 1 GeV DM, resulting in a larger value of $\La_3$
and thus a smaller photon flux. For 1 GeV DM, the photon flux is comparable to the GC $\ga$-ray excess.
Note that the values of $\Lambda_3$ assumed here actually lead to a smaller DM annihilation cross-section than required, i.e., one will end up with a too large DM relic density. This issue can be solved by
including $\Lambda_1$ and $\Lambda_2$ terms in Eq.~\eqref{eq:Lagr} to increase the DM annihilation rate during freeze-out, but the polarization rate will be reduced as the $\ga_5$ terms have $\vec{L}=\vec{S}=0$ contributions.

To conclude, the possibility of detecting polarized photons will depend on how many of DM particles at the GC are polarized and also depend on the DM mass. For a light DM below 5 GeV, a preferred direction for the photon polarization could potentially be detectable if a significant part of the DM particles is polarized along a certain direction around the GC.
For heavier DM, the photon flux is hopelessly small because of the stringent direct search bounds.

\section{Conclusions and outlook} \label{section:conclusion}

The azimuthal angle of the plane of production of an electron-positron pair created in a $\ga$-ray detector provides a way of measuring linear polarization of incoming $\ga$ rays. The current $\ga$-ray detectors are not designed primarily for polarization measurement. Instruments sensitive to linear polarization will be employed in future $\ga$-ray experiments such as AdEPT and ASTROGAM, with the minimum detectable polarization (MDP) from a few percents up to $20\%$ or so~\cite{Knodlseder:2016pey}. 

In  this work, we have proposed a simple phenomenological model
where two types of Majorana DM particles, degenerate in mass, have anapole interactions. 
The anapole interactions can polarize the spins of DM in the presence of electric currents at the GC, and 
in turn the linearly polarized $\ga$-ray flux from DM annihilations can be realized.
The degree of polarization of the $\ga$ rays can reach as much as $70\%$ at the DM mass threshold,
given the mass ratio of the final state fermion to DM being 0.1. For a larger mass ratio, a higher polarization rate is expected.
 For DM mass of about $1$ GeV, the $\ga$-ray flux can be comparable to the $\ga$-ray excess in the GC. The origin of the GC excess is still a puzzle, probably comprised of diffuse $\ga$ rays from multiple components of many different sources and thus being most likely unpolarized. Therefore, any detection of polarized $\ga$ rays may give an invaluable understanding of the GC excess. Even though the DM induced $\ga$-ray flux is an order of magnitude below the GC excess, the polarization measurement can still be used to identify a highly polarized $\ga$-ray signal. In addition, the photon flux induced by GeV DM has a plateau shape extending to sub-GeV energies, though the degree of polarization drops to $\lesssim 10\%$. This polarized low-energy $\ga$-ray signal may be of interest to current $X$-ray telescopes such as the POLAR satellite, which is a X-ray polarimeter sensitive to an energy up to $0.5$ MeV with a $10\%$ MDP~\cite{Kole:2018hbo}. In addition, the Advanced Energetic Pair Telescope (AdEPT~\cite{Hunter:2013wla}) and the time projection chamber as a gamma-ray telescope and polarimeter~(HARPO~\cite{Gros:2017wyj}) have been proposed to study the photon polarization at few percent level for the sub-GeV energy range.

As we have explained how the spins of anapole DM can be aligned with the electric current flow in the GC, unfortunately it is rather difficult to determine the degree of alignment without knowing the details of the formation of the dark halo and the accretion disk. A low degree of alignment will definitely degrade the detectability of a polarized signal. Here we stress that the present paper has given a first attempt to investigate a possible {\it linearly} polarized $\ga$-ray signal from DM annihilations. It is certainly important to explore further along this direction for making full use of the polarization capability of future $\ga$-ray detectors.


\section*{Acknowledgments}

WCH is grateful for the hospitality of IOP Academia Sinica and NCTS in Taiwan,
where this work was initiated.
WCH was supported by DGF Grant No. PA 803/10-1 and by the Independent Research Fund Denmark, grant number 
DFF 6108-00623.
KWN is supported by Ministry of Science and Technology, Taiwan, ROC under the Grant
No. MOST104-2112-M-001-039-MY3.
The CP3-Origins centre is partially funded by the Danish National Research Foundation, grant number DNRF90.



\end{document}